# Superconductivity induced by doping holes in the nodal-line semimetal NaAlGe

Toshiya Ikenobe[1], Takahiro Yamada[2], Daigorou Hirai[1*], Hisanori Yamane[2], and Zenji Hiroi[1]

[1] *Institute for Solid State Physics, University of Tokyo, Kashiwa, Chiba 277-8581, Japan*

[2] *Institute of Multidisciplinary Research for Advanced Materials, Tohoku University, Katahira 2-1-1, Aoba-ku, Sendai 980-8577, Japan*

\* *Present address: Department of Applied Physics, Nagoya University, Nagoya 464–8603, Japan*

The nodal-line semimetals NaAlSi and NaAlGe have significantly different ground states despite having similar electronic structures: NaAlSi exhibits superconductivity below 7 K, while NaAlGe exhibits semiconductive electrical conductivity at low temperatures, indicating the formation of a pseudogap at approximately 100 K. The origin of the pseudogap in NaAlGe is unknown but may be associated with excitonic instability. We investigated hole-doping effects on the ground state in the solid solution Na(Al$_{1-x}$Zn$_x$)Ge and discovered that the pseudogap is suppressed continuously with increasing Zn content, followed by the appearance of a superconducting dome with the highest transition temperature of 2.8 K. This superconductivity most likely results from excitonic fluctuations.

## I. INTRODUCTION

Topological semimetals with Dirac points close to the Fermi level have garnered significant attention because of their unique transport properties [1-3]. When multiple Dirac points form a line in momentum space, the system is referred to as a nodal-line semimetal, which has been focused theoretically [4,5] and studied in materials such as CaAgP [6,7], β-ReO$_2$ [8,9], and PbTaSe$_2$ [10-12]. Recent research has focused on electronic instabilities caused on by perturbations such as electron correlations in topological semimetals [13-17]. In nodal-line semimetals, it is expected that electron correlations will induce various types of instability leading to charge-density-wave, antiferromagnetic, excitonic, or superconducting phases [18-23]. This article highlights the fascinating physical properties of the nodal-line semimetal NaAlGe.

Both NaAlGe and the related compound NaAlSi crystallize in anti-PbFCl type layered structures (space group *P*4/*nmm*) [24]. In the structures, Al-centered Ge (Si) tetrahedra are connected by edge-sharing to form conducting layers that alternate along the *c*-axis with block layers composed of double Na sheets [25]. First-principles electronic state calculations indicate that both compounds have similar electronic structures, characterized by the highly dispersive electron-like Al-3*s* band and the less dispersive hole-like Ge-4*p* (Si-3*p*) band, which cross each other to form a nodal line near the Fermi energy $E_F$ [26-29]. In the presence of spin–orbit interaction (SOI), however, the band crossing points are eliminated, and small gaps of 30 K and 10 K on average open for NaAlGe and NaAlSi, respectively.

Despite their similar crystal and electronic structures, the ground states of NaAlGe and NaAlSi are distinct. NaAlSi exhibits metallic behavior over a broad temperature range and superconductivity at the critical temperature $T_c \sim 7$ K [30, 31]. This is conventional *s*-wave superconductivity mediated by electron–phonon interactions [28,31-33]. NaAlGe, in contrast, exhibits a semiconducting increase in electrical resistivity and a decrease in magnetic susceptibility and carrier density below 100 K [34]. These findings unequivocally indicate a decrease in the density of states (DOS), suggesting the formation of a pseudogap. The origin of the pseudogap has been proposed as an excitonic instability, which may be related to correlation effects in the nodal-line band [18,19]. Why the two compounds have such different ground states despite their apparent similarities is an intriguing question.

NaAlSi and NaAlGe are both chemically unstable in the atmosphere. In particular, the latter reacts with atmospheric moisture and degrades rapidly [34]. Chen et al. recently reported that NaAlGe exhibited superconductivity at temperatures below 3 K during degradation prior to complete decomposition [35]. It was argued that the appearance of superconductivity was the result of hole doping caused by the degradation-induced loss of Na in a portion of the crystal. However, detailed characterization of superconducting properties was challenging because the sample obtained was heterogeneous and changed as degradation progressed.

In order to investigate the possibility of superconductivity in NaAlGe, we have systematically examined hole-doping effects by substituting Zn for Al in a series of polycrystalline samples of Na(Al$_{1-x}$Zn$_x$)Ge. Electrical resistivity, magnetic susceptibility, and heat capacity measurements reveal that bulk superconductivity appears at 1.5–4.5% substitution, with the highest $T_c$ of 2.8 K occurring at approximately 3% substitution, resulting in a superconducting $T_c$ dome. Moreover, the pseudogap is suppressed by hole doping and tends to disappear near the dome's apex. The superconductivity of Na(Al$_{1-x}$Zn$_x$)Ge is likely attributable to fluctuations associated with the pseudogap formation, such as excitonic fluctuations.

## II. EXPERIMENTAL
### A. Sample preparation

We synthesized polycrystalline samples of Na(Al$_{1-x}$Zn$_x$)Ge using as starting materials Na lump (Nippon Soda, 99.95%), Al rod (5mm diameter; Nilaco, 99.9999%), Ge lump (Kojundo Chemical Laboratory, 99.9999%), and Zn shot (Strem Chemicals, Inc., 99.99%). Na : Al : Zn : Ge = 1 : 1 − $x_n$ : $x_n$ : 1 (0 < $x_n$ < 0.15) were weighed (total weight was about 400 mg) and placed in a BN crucible sealed in a stainless steel container. All operations were performed in a glove box filled with Ar gas to prevent deterioration caused by reaction with moisture in the air. The container was heated at 1123 K for 4 hours in an electric furnace. The obtained product was crushed, mixed, and pressed into a 3 × 3 × 14 mm$^3$ pellet before being re-heated at 998 K for 40 hours in a BN crucible and placed in a stainless-steel container. The final product was obtained after heating at 998 K for 100 hours in the same manner.

The chemical composition of the samples was determined using wavelength-dispersive X-ray (WDX) spectroscopy in an electron probe microanalyzer system (JEOL XA-8200). No notable inhomogeneous distribution of composition was detected. The average chemical composition of the $x_n$ = 0 sample was Na$_{1.09}$Al$_{1.00}$Ge$_{0.91}$, indicating that the Ge site was replaced by approximately 10% more Na; similar results were reported for single crystals prepared using the Na–Ga flux method, such as $Na_{1.13}Al_{0.97}Ga_{0.01}Ge_{0.89}$ [34]. How this exchange of Na and Ge affects the Fermi level is not trivial. Our sample exhibited a semiconductor-like conductivity comparable to that observed in earlier single crystals [34]; thus, the modification may not be crucial for the issue of interest. This type of compositional deviation was always observed in Zn-substituted samples, regardless of $x_n$. The actual Zn content $x$ determined by WDX measurements was smaller than the nominal composition $x_n$. As shown in Fig. 1(a), $x_n$ and $x$ have a linear relationship with a slope of 0.35(1). Consequently, approximately one-third of the nominal Zn content remains in the sample. From now on, we will use $x$ instead of $x_n$.

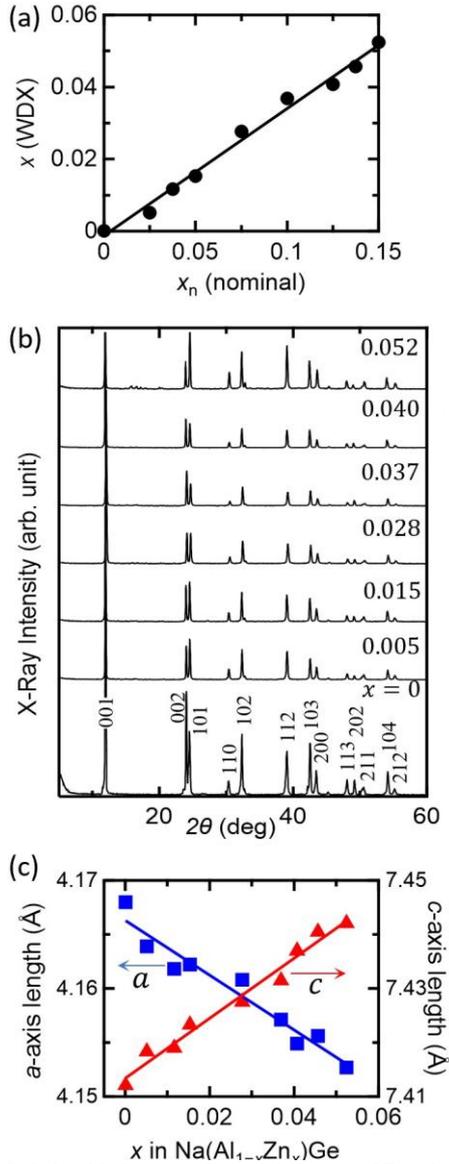

FIG. 1. (a) Relationship between the actual amount of Zn substitution $x$ as determined by WDX measurements and the nominal substitution $x_n$ in preparations. The slope of the solid line is 0.35(1). (b) Powder XRD patterns of $Na(Al_{1-x}Zn_x)Ge$. (c) Composition dependences of the lattice constants. The solid lines are linear fits to the data points: $a = -0.25(2)x + 4.166(1)$ and $c = 0.56(3)x + 7.413(1)$.

Powder X-ray diffraction (XRD) measurements using a diffractometer (Bruker, D2 PHASER, Cu-K$\alpha$) under an Ar atmosphere were used to characterize the obtained samples. The powder XRD patterns of Fig. 1(b) show that all the samples are nearly monophasic. The lattice parameters were calculated by the Le Bail method using the TOPAS (Bruker) software. The $x = 0$ sample has lattice constants of $a = 4.1680(2)$ Å and $c = 7.4121(3)$ Å, which are close to the previously reported values of $a = 4.1634(2)$ Å and $c = 7.4146(4)$ Å for a NaAlGe single crystal [34]. The composition dependences are shown in Fig. 1(c). As $x$ increases, $a$ decreases and $c$ increases linearly, in accordance with Vegard's law. As a result, samples with systematic Zn substitution in the range of 0–5.2% were successfully obtained.

B. Characterizations

A physical property measurement system (PPMS, Quantum Design) was used to measure electrical resistivity and heat capacity. Electrical resistivity was measured using a four-terminal method with indium terminals crimped onto the crystal surface; silver paste terminals could not be used due to their high contact resistance. Before transferring the sample with electrodes from the glove box to the sample chamber of PPMS, it was necessary to cover the sample with liquid paraffin to prevent degradation by moisture in the air. As shown in Fig. 2, the measured electrical resistivity of the $x = 0$ sample exhibited a semiconductive temperature dependence comparable to the previous data from single crystals. In contrast, when the electrical resistivity was measured without a liquid paraffin cover, the absolute value was reduced by one order of magnitude, and the increase below 100 K became weaker, indicating the introduction of additional carriers. Moreover, a slight decrease in resistivity occurred around 2 K, which may be the result of degradation-induced partial superconductivity as previously reported [35].

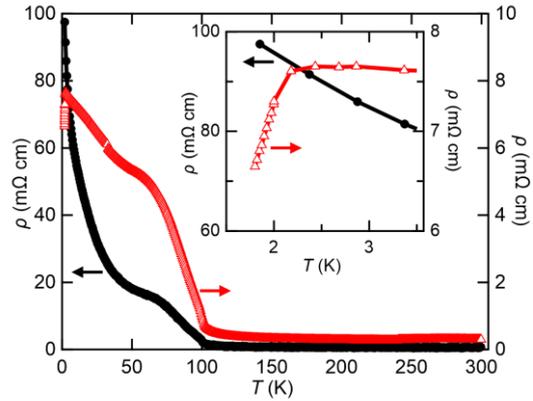

FIG. 2. Temperature dependence of electrical resistivity for two samples of NaAlGe. The black filled circles represent data for a sample covered with liquid paraffin (left axis), whereas the red open triangles represent data for a sample without liquid paraffin (right axis). The inset shows a magnified view of the low-temperature region, where a slight decrease due to partial superconductivity is likely observed only for the sample without paraffin coating.

Heat capacity measurements were carried out in a PPMS via the relaxation method. Magnetic susceptibility was measured in a magnetic property measurement system (MPMS, Quantum Design). To minimize degradation, a sample pellet was coated with grease in a glove box and inserted into the sample chamber filled with He gas.

## III. RESULTS
### A. Superconductivity

Figure 3(a) depicts the temperature dependences of electrical resistivity below 3.5 K for samples substituted with 0.5%–5.2% Zn. The electrical resistivity data of the 0.5% and 1.0% samples are almost independent of temperature down to 1.8 K, while the resistivity of the 1.5% sample drops from 2.5 K to zero below 2.0 K. The superconducting transition temperature $T_c$ is 2.3 K, as defined by the temperature at which the linear extrapolation of the drop crosses zero. As doping increases, the transition temperature rises, with a maximum $T_c$ of 2.8 K for the 2.8% sample. Subsequently, $T_c$ decreases to 2.5 K (3.7%), 2.2 K (4.0%), and 2.0 K (4.6%), and in the 5.2% substituted sample, the transition begins at approximately 1.8 K; $T_c < 1.8$ K.

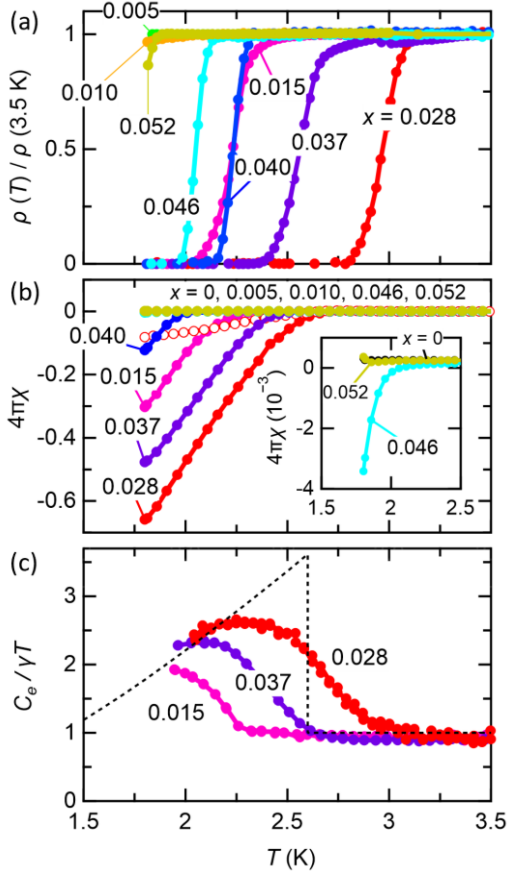

FIG. 3. (a) Evolution of the temperature-dependent electrical resistivity $\rho$, normalized to $T = 3.5$ K, as a function of the Zn content. (b) Temperature dependences of magnetic susceptibility $\chi$. The majority of data curves depicted by filled circles were recorded in a magnetic field of $H = 5$ Oe after cooling to 1.8 K in zero magnetic field, while the red open circles depicted field-cooled data for the 2.8% sample in the same field. The inset magnifies the datasets below 2.5 K. (c) Temperature dependence of $C_e/\gamma T$ for the 1.5%, 2.8%, and 3.7% substituted samples, where $C_e$ and $\gamma$ are, respectively, the electronic part of the heat capacity and the Sommerfeld coefficient. The broken line shows the approximate, ideal variation expected for the 2.8% curve, taking into account transition broadening and entropy balance.

The magnetic susceptibility and heat capacity data shown in Fig. 3(b) and (c) indicate that the bulk nature of the superconductivity, as opposed to a partial character associated with impurities or inhomogeneities [35], is evident. For the 2.8% sample, the magnetic shielding and Meissner volume fractions reach 65% and 10% at 1.8 K, respectively, which are large enough to insist bulk superconductivity; the former is less than 100% due to magnetic field penetration in the polycrystalline sample, while the latter is smaller due to magnetic flux pinning effects. The 0%, 0.5% and 1.0% samples do not exhibit diamagnetic response above 1.8 K, whereas the 1.5% sample does below 2.4 K. As $x$ increases, $T_c$, which is the onset temperature of diamagnetic response, reaches a maximum of 2.8 K for the 2.8% sample, then decreases to 2.6 K (3.7%), 2.4 K (4.0%), 2.0 K (4.6%), and finally below 1.8 K for the 5.2% sample, which does not exhibit diamagnetic response above this temperature. Thus, the variations of $T_c$ in magnetic susceptibility are in excellent agreement with variations in electrical resistivity.

Figure 3(c) illustrates the temperature dependences of $C_e/\gamma T$ for 1.5%, 2.8%, and 3.7% samples, for which the electronic heat capacity $C_e$ and the Sommerfeld coefficient $\gamma$ were estimated from the $C/T$ versus $T^2$ plot (Fig. 8). For each composition, a characteristically large peak in heat capacity is observed, which is indicative of a second-order phase transition. The transitions are broadened relative to the ideal BCS curve, such as the broken curve for the 2.8% sample, which is depicted approximately assuming the entropy balance. The cause of the broadening must be sample inhomogeneity and randomness effects, which are unavoidable for such solid solution systems. $T_c$s are defined as 2.1 K (1.5%), 2.6 K (2.8%), and 2.3 K (3.7%) at the midpoint of the jump. For the 2.8% sample, the magnitude of the increase across the transition $\Delta C_e/\gamma T$ is 1.6 for the experimental curve and greater than 2 for the curve considered entropy balance. These values are significantly larger than 1.43 expected for weak-coupling BCS superconductivity, indicating the realization of strong-coupling superconductivity.

All measurements of electrical resistivity, magnetic susceptibility, and heat capacity indicate that hole doping with the Zn-for-Al substitution transforms NaAlGe into a superconductor. The $T_c$ values measured by the three experimental probes are consistent with one another. Figure 4 provides a summary of $T_c$'s doping dependences: $T_c$ could be lower than our experimental limit of 1.8 K at 0.5% and 1.0 %, rises to 2.2–2.3 K at 1.5%, reaches a maximum of 2.8 K at 2.8%, decreases to 2.0 K at 4.6%, and falls below 1.8 K at 5.2%. Through systematic hole doping with Zn substitution, a dome-shaped superconducting phase appears. Chen et al.'s observation of superconductivity with $T_c$ of 3 K in degraded NaAlGe [35] may correspond to our optimal doping concentration of 2.8 %.

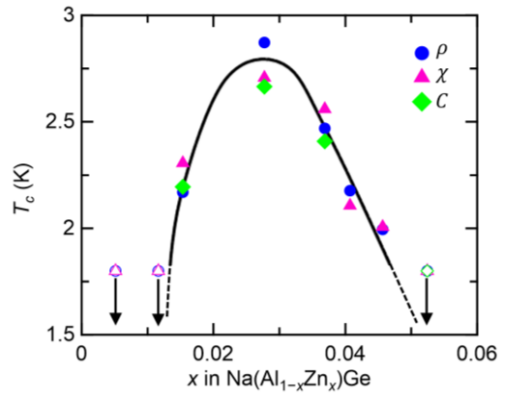

FIG. 4. Hole doping dependence of $T_c$ derived from electrical resistivity (blue circles), magnetic susceptibility (magenta triangles), and heat capacity (green diamonds) measurements (Fig. 3). The samples for which superconductivity was not observed above 1.8 K are represented by open symbols with arrows pointing downward. The solid lines serve as visual aids.

## B. Normal-state properties

Here, the doping effects on the normal-state properties of NaAlGe are described. Figure 5(a) illustrates the dependence of electrical resistivity on doping below 300 K. The electrical resistivity of the pure sample exhibits the same temperature dependence as the previous single crystal: metallic behavior at room temperature, followed by an increase below ~100 K [34]. However, the increase is saturated at low temperatures, indicating that NaAlGe is not a simple insulator but a metal with a pseudogap. Doping diminishes the increase in resistivity, and shifts it to lower temperatures, reaching approximately 50 K and 20 K for 1.0% and 1.5% dopings, respectively. Above 2.8%, the resistivity decreases monotonically and exhibits typical metallic behavior. Thus, the pseudogap is suppressed by hole doping.

The pseudogap temperature $T^*(\rho)$ can be approximated as the temperature at which the effective activation energy $E_\rho$, defined as $k_B T^2(\partial \ln\rho/\partial T)$, reaches its maximum [36-38]. As shown in Fig. 5(b), the $E_\rho$ of the pure sample exhibits a clear peak at $T^*(\rho) = 91$ K. $T^*(\rho)$ decreases systematically with increasing doping: 58 K (0.5%), 39 K (1.0%), and 29 K (1.5%). Then, there is no peak above 2.8%.

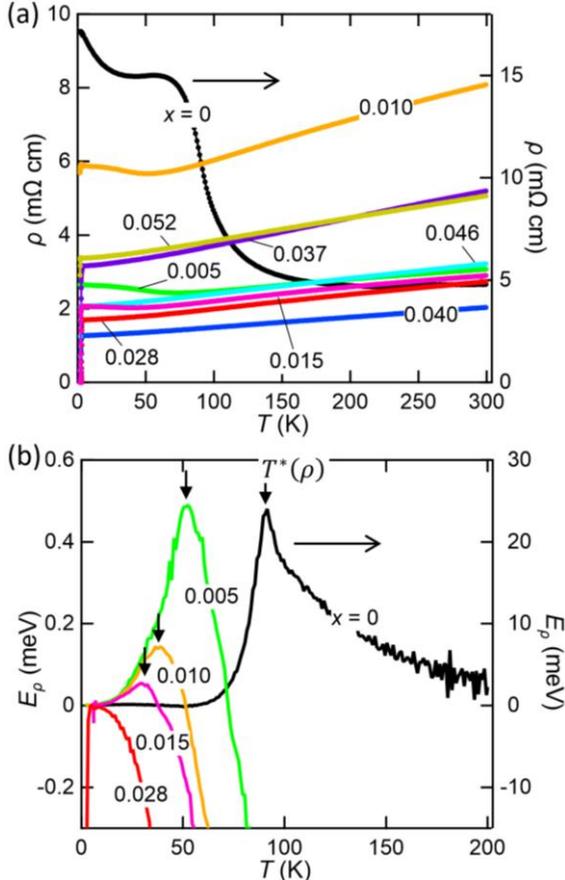

FIG. 5. (a) Temperature dependence of electrical resistivity for all composition samples from 2 to 300 K. (b) Temperature dependence of $E_\rho$, defined as $k_B T^2(\partial \ln\rho/\partial T)$, for 0–2.8% substituted samples. The temperatures depicted by the arrows correspond to the pseudogap temperature $T^*(\rho)$. In both (a) and (b), only the data from the pure sample are plotted along the right vertical axis.

The temperature dependence of magnetic susceptibility is depicted in Fig. 6(a). The magnetic susceptibility of the pure sample decreases significantly below 100 K, as observed for the previous single crystal, and this decrease is associated with the formation of a pseudogap [34]. With doping, the decrease becomes less pronounced, shifts to lower temperatures, and is no longer observed for substitution rates above 2.8%. The pseudogap temperature $T^*(\chi)$ is defined as the maximum temperature of the temperature derivative $d\chi/dT$ [Fig. 6(b)]. $T^*(\chi)$ decreases as doping increases: 98 K (0%), 44 K (0.5%), 32 K (1.0%), and 24 K (1.5%).

The magnetic susceptibility of the NaAlGe sample at 2 K is $6.1 \times 10^{-5}$ cm$^3$ mol$^{-1}$, which is nearly identical to the previous single crystal's $6.2 \times 10^{-5}$ cm$^3$ mol$^{-1}$ [34]. The magnitude decreases dramatically with doping, falling by less than half to $2.8 \times 10^{-5}$ cm$^3$ mol$^{-1}$ at 5.2%, which approach calculated value $\chi_{\text{band}} = 1.9 \times 10^{-5}$ cm$^3$ mol$^{-1}$. This anomalous decrease may be crucial to comprehending the change in the electronic state, and it will be discussed later.

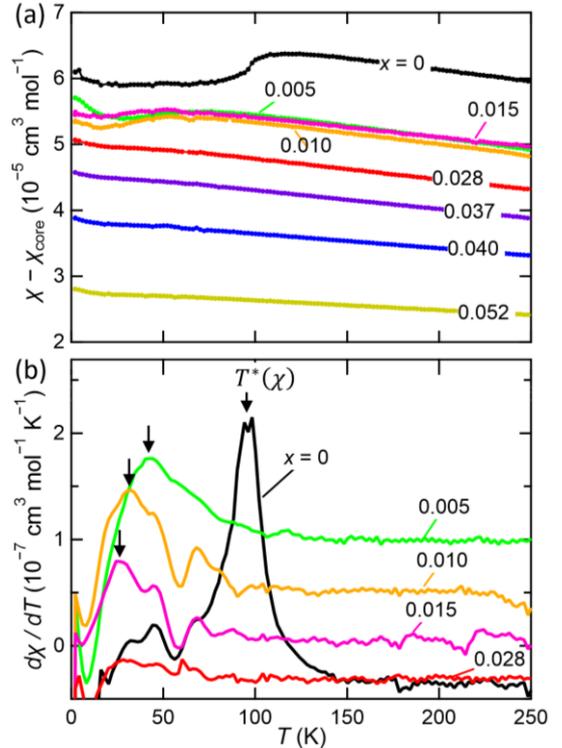

FIG. 6. (a) Temperature dependence of magnetic susceptibility measured at $B = 7$ T after the subtraction of nuclear diamagnetic contributions for samples of various compositions. (b) Temperature derivative of magnetic susceptibility $d\chi/dT$ for 0–2.8% substituted samples. Offsets were added for clarification. The maximum temperature indicated by the arrow is the pseudogap temperature $T^*(\chi)$.

## C. Phase diagram

The superconducting transition temperature $T_c$ and pseudogap formation temperature $T^*$ are summarized in the $T$–$x$ phase diagram of Fig. 7. These temperatures obtained by measuring electrical resistivity, magnetic susceptibility, and heat capacity measurements, are almost identical. $T^*$ decrease rapidly upon doping, with the pseudogap phase disappearing at approximately 2%. The superconducting phase appears between 1.5% and 4.6% above 1.8 K, with the highest $T_c \sim 2.8$ K occurring at 2.8%. The results indicate that a dome-shaped superconducting phase appears in Na(Al$_{1-x}$Zn$_x$)Ge following the suppression of the pseudogap phase.

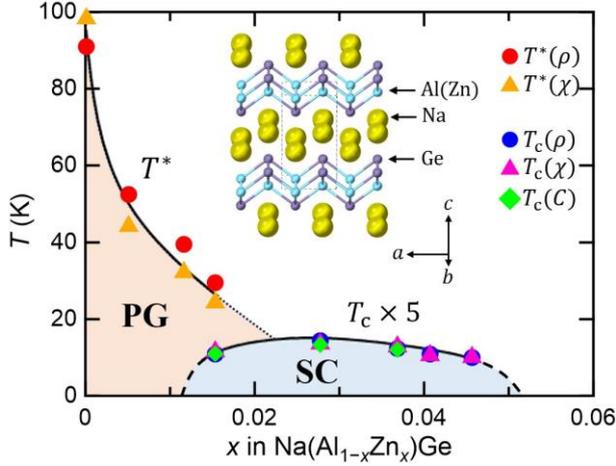

FIG. 7 $T$–$x$ phase diagram of Na(Al$_{1-x}$Zn$_x$)Ge. $T^*$ and $T_c$ (multiplied by a factor of 5) from electrical resistivity (circles), magnetic susceptibility (triangles), and heat capacity measurements (diamonds) are plotted as a function of $x$. The pseudogap and superconducting phases are denoted by PG and SC, respectively.

### D. Doping dependence of the electronic state

Figure 8 depicts the low-temperature heat capacity of the 0%, 1.5%, 3.7%, and 5.2% polycrystalline samples, as well as that of the 0% single crystal sample reported in the literature [34]. The typical relationship $C(T) = \gamma T + \beta T^3$ holds below 4 K, and the Sommerfeld coefficient $\gamma$ is derived from the $C/T$ versus $T^2$ intercept. The experimental values of $\gamma_{exp}$ are 0.45(2) and 0.45(1) mJ K$^{-2}$ mol$^{-1}$ for the single crystal and polycrystalline 0% samples, respectively, which are comparable values. It gradually increases with doping to 1.55(3) mJ K$^{-2}$ mol$^{-1}$ for the 5.2% sample (Fig. 9). On the other hand, doping increases the slope $\beta$. The Debye temperature $\Theta_D$ calculated using the formula $\Theta_D = ((12\pi^4 NR)/5\beta)^{1/3}$ ($N$ is the number of atoms per formula unit, and $R$ is the gas constant) are 235 (0% single crystal), 213 (0%), 206 (1.5%), 182 (3.7%), and 192 K (5.2%); the difference between the polycrystalline and single crystal values for the 0% sample may reflect the polycrystalline nature. As a result, the lattice tends to soften as Zn substitution increases.

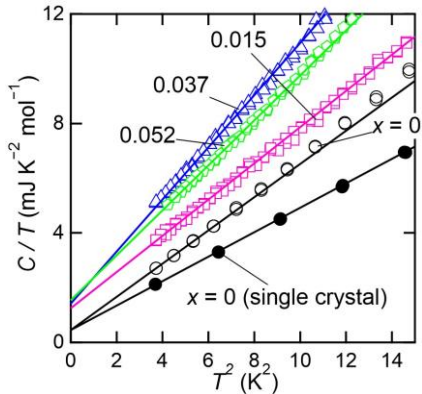

FIG. 8. $C/T$ versus $T^2$ plots for 0% (black circles), 1.5% (magenta squares), 3.7% (blue triangles), and 5.2% (green pentagons) polycrystalline samples, and 0% single crystal (filled black circles) [34]. The straight lines are fits to the equation $C(T) = \gamma T + \beta T^3$.

Figure 9 illustrates the Zn-substitution dependences of the magnetic susceptibility at 2 K, the Sommerfeld coefficient, and the Wilson ratio $R_W$. The $\gamma_{exp}$ of the parent phase is much smaller than the calculated $\gamma_{band}$ of 1.4 mJ K$^{-2}$ mol$^{-1}$, which is attributed to the reduction of DOS due to the formation of pseudogap [34]; in NaAlSi, $\gamma_{exp}$ and $\gamma_{band}$ are 2.15 and 1.7 mJ K$^{-2}$ mol$^{-1}$, respectively, and there is rather a 30% enhancement in the experimental value [31], which may be due to an electron–phonon interaction. Doping causes $\gamma_{exp}$ to increase and saturate close to the $\gamma_{band}$, indicating suppression of the pseudogap.

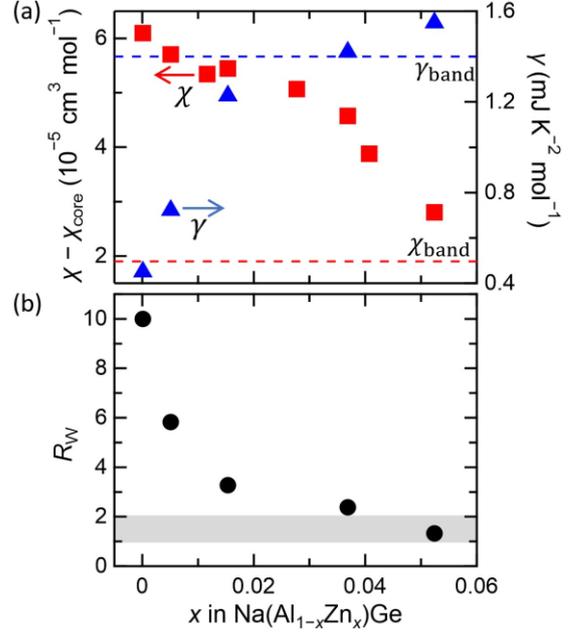

FIG. 9. (a) Doping dependences of the magnetic susceptibility at $T$ = 2 K after subtraction of the nuclear diamagnetic contribution $\chi_{core}$ and the Sommerfeld coefficient $\gamma_{exp}$ on the left vertical and right vertical axes, respectively. The $\chi_{band}$ and $\gamma_{band}$ obtained from the band calculations are shown by the red and blue dotted lines, respectively. (b) Doping dependence of the Wilson ratio $R_W$.

In stark contrast to the dependence of the Sommerfeld coefficient on doping, the magnetic susceptibility decreases gradually at first, then rapidly around 4%, and reaches approximately half at 5.2%. In comparison to the calculated value of $1.9 \times 10^{-5}$ cm$^3$ mol$^{-1}$ [34], the values at 0% and 5.2% are 3.2 and 1.5 times larger, respectively, and the enhancement decreases with increasing doping. The behavior observed in Fig. 9(a) is atypical given that both $\gamma$ and $\chi$ are proportional to DOS in typical Pauli paramagnetic metals, where the ratio of the two, the Wilson ratio $R_W = (\pi^{2/3})(k_B/\mu_0)^2(\chi/\gamma)$, is 1 for free electron systems and 2 for strongly correlated electron systems [39]. The doping dependence of the $R_W$, as depicted in Fig. 9(b), reveals that the anomalously large $R_W = 10$ (0%) decreases to 5.8 (0.5%), 3.3 (1.5%), 2.4 (3.7%), and 1.3 (5.2%), thus asymptotically approaching the free electron model value. Interestingly, the $R_W$ of NaAlSi is 2.3, which is significantly lower than that of NaAlGe but still greater than the value predicted by the free electron model [31]. Note that typical electron–phonon interactions increase $\gamma$ while maintaining $\chi$.

Consequently, NaAlGe has an anomalous electronic state with a large enhancement in magnetic susceptibility and a large reduction in the Sommerfeld coefficient, both of which are eliminated by hole doping. Given that NaAlGe is an $sp$-electron system, conventional electronic correlations must be small, and magnetic instability has not been observed. In addition, there are no Fermi surface instabilities such as nesting, and there may be no structural instability. Therefore, there must be a unique component causing the unusual electronic state. On the other hand, the moderate enhancement of magnetic susceptibility in NaAlSi may have a common origin. It is likely that their common nodal lines are responsible for the enhanced magnetic susceptibility, which is especially pronounced in NaAlGe.

## IV. DISCUSSION

The superconducting dome observed in Na(Al$_{1-x}$Zn$_x$)Ge is typical for carrier-doped superconductors such as the copper, iron and $f$ electron-based superconductors [40-42]. It is commonly believed that the mechanism of superconductivity is related to fluctuations caused by the suppression of a certain order in the parent phase by doping; in many compounds, this order is magnetic or structural. In contrast, doping suppresses the pseudogap phase in Na(Al$_{1-x}$Zn$_x$)Ge, and the superconducting dome appears. If the origin of the pseudogap in NaAlGe is due to excitonic instability, the superconductivity in Na(Al$_{1-x}$Zn$_x$)Ge could be induced by excitonic fluctuations. In the case of the other nodal-line semimetal ZrSiS, the formation of a pseudogap due to excitonic instability has been postulated, and it has been pointed out that doping strongly suppresses the pseudogap by breaking the electron-hole symmetry, leading to $d$-wave superconductivity [19,43].

A few excitonic insulator candidates claim to exhibit superconductivity. In Ta$_2$NiSe$_5$, for instance, it was reported that superconductivity emerged under high pressure [44]. However, it is unclear whether the superconductivity is associated with excitonic fluctuations, as it occurs in a crystal structure distinct from the excitonic insulator phase at low pressures. In ZrSiS [18], on the other hand, tip-induced superconductivity was observed, but its superconductivity was not well understood [45]. Accordingly, to the best of our knowledge, bulk excitonic fluctuation-induced superconductivity has not been demonstrated in any materials. It would be quite interesting if Na(Al$_{1-x}$Zn$_x$)Ge were the first excitonic fluctuation-induced superconductor. Future experiments will require angle-resolved photoemission spectroscopy or tunneling electron spectroscopy to investigate the superconducting gap structure using a single crystal Na(Al$_{1-x}$Zn$_x$)Ge.

Considering that NaAlSi is a superconductor due to electron–phonon interaction [28,31], a simple electron–phonon interaction is a potential alternative superconducting mechanism. The pseudogap and superconducting states in the phase diagram may be considered to be in competition with one another, as opposed to cooperating as expected in the previously described excitonic scenario. Doping may suppress the excitonic instability in NaAlGe and replace it with electron–phonon instability, resulting in electron–phonon superconductivity; the excitonic instability, which destroys superconductivity, has been removed by doping. The Zn substitution may enhance electron–phonon interactions by decreasing the Debye temperature. In previous experiments on NaAl(Si$_{1-x}$Ge$_x$), $T_c$ decreased with increasing Ge substitution, and bulk superconductivity disappeared above 45% [46], indicating that electron–phonon interactions were weakened on the Ge-rich side. Thus, an electron–phonon scenario cannot be ruled out in Na(Al$_{1-x}$Zn$_x$)Ge, as the Si and Zn substitutions restore the weakened electron–phonon interaction in NaAlGe. Note, however, that the hole-doping-induced superconductivity lies in the strong-coupling regime, as evidenced by the large $\Delta C_e/\gamma T$ greater than 2 (Fig. 3), whereas the superconductivity of NaAlSi lies in the weak-coupling regime despite its higher $T_c$ [31]. This comparison of superconductivity property suggests the existence of distinct mechanisms. In any case, the anomalous pseudogap state with large $R_W$ of NaAlGe is remarkable, and the resulting superconductivity must be fascinating. We believe that interesting physics related to the electronic instability of the nodal-line semimetal have yet to be discovered.

## V. SUMMARY

The nodal-line semimetal NaAlGe has an unusual electronic structure characterized by the formation of a pseudogap and an anomalously large Wilson ratio, which may be indicative of excitonic instability. We carried out hole doping by Zn-for-Al substitution and discovered a dome-shaped superconducting phase with the highest $T_c$ of 2.8 K in the 1.5%–4.6% substituted samples. Furthermore, the pseudogap phase is continuously suppressed and replaced by the superconducting phase as hole doping increases. It is likely that excitonic fluctuations are responsible for the superconductivity of Na(Al$_{1-x}$Zn$_x$)Ge.

### Acknowledgments

The authors are grateful to Yoshihiko Okamoto for insightful discussion. This research was financially supported by JSPS KAKENHI Grants (JP20H02820, JP20H01858, JP22H04462 and JP22H05147) and Cooperative Research Program of "Network Joint Research Center for Materials and Devices" (20225008). The authors would like to thank Chikako Nagahama for her help in the sample preparation.